\documentclass[letterpaper]{article} 
\usepackage{aaai2026}  
\usepackage{times}  
\usepackage{helvet}  
\usepackage{courier}  
\usepackage[hyphens]{url}  
\usepackage{graphicx} 
\usepackage{natbib}  
\usepackage{caption} 
\usepackage{booktabs} 
\usepackage{algorithm}
\usepackage{algorithmic}

\usepackage{tcolorbox}
\usepackage{subcaption}

\usepackage{multirow}

\usepackage{newfloat}
\usepackage{listings}
\DeclareCaptionStyle{ruled}{labelfont=normalfont,labelsep=colon,strut=off} 
\floatstyle{ruled}
\newfloat{listing}{tb}{lst}{}
\floatname{listing}{Listing}
\usepackage{xcolor}

\usepackage{newfloat}
\usepackage{listings}
\DeclareCaptionStyle{ruled}{labelfont=normalfont,labelsep=colon,strut=off} 
\floatstyle{ruled}
\newfloat{listing}{tb}{lst}{}
\floatname{listing}{Listing}
\title{The Trafficker’s Pitch: Detecting Deceptive Recruitment in Online Job Boards}
\author{
    Siyi Zhou \textsuperscript{\rm 1, 3} \thanks{}, Peiran Qiu \textsuperscript{\rm 2}, Tanishq Salkar\textsuperscript{\rm 2,3}, Leonardo Blas Urrutia \textsuperscript{\rm 2,3}, Dacheng Shen\textsuperscript{\rm 2}, Nora Adadurova \textsuperscript{\rm 4} , Deyang Hsu \textsuperscript{\rm 2} , Eun Cheol Choi \textsuperscript{\rm 2, 1}, Emilio Ferrara \textsuperscript{\rm 1, 2, 3}
}
\affiliations{
    \textsuperscript{\rm 1}Annenberg School of Communication and Journalism,\\
    \textsuperscript{\rm 2} Thomas Lord Department of Computer Science, School of Advanced Computing,\\

    \textsuperscript{\rm 3}Information Science Institute,\\
    University of Southern California
    Los Angeles, CA, USA\\
    \textsuperscript{\rm 4}San Jose State University

    \{zhousiyi,peiranqi, salkar, blasurru, dachengs, deyanghs, euncheol, emiliofe\}@usc.edu
}

\usepackage{bibentry}

\begin{document}

\begin{figure*}[t]
\centering
\begin{tcolorbox}[colback=white, colframe=red, boxrule=1.2pt, arc=0pt,
                   left=6pt, right=6pt, top=6pt, bottom=6pt, width=\textwidth]
\centering
\textbf{Erratum:} The version of this paper published in the Proceedings of ICWSM 2026 (CySoc Workshop) omitted an author due to a technical error during submission. This version contains the correct, complete author list.

\vspace{4pt}
\textbf{Please cite this corrected version as:}\\
Zhou, S., Qiu, P., Salkar, T., Blas Urrutia, L., Shen, D., Adadurova, N., Hsu, D., Choi, E.~C., \& Ferrara, E. (2026). The Trafficker's Pitch: Detecting Deceptive Recruitment in Online Job Boards. In \textit{Proceedings of the International AAAI Conference on Web and Social Media (ICWSM), CySoc Workshop}.
\end{tcolorbox}
\end{figure*}
\maketitle

\begin{abstract}
While substantial efforts in anti-trafficking research and practice have focused on identifying and assisting victims after exploitation occurs, comparatively less attention has been paid to preventing victimization at the recruitment stage. Although some platforms offer preventive tools, such as background checks triggered by in-person meeting detection, these measures primarily protect potential victims rather than directly limiting traffickers’ recruitment activities. In this paper, we propose a computational framework to identify human trafficking recruiters through their linguistic features and to characterize their online recruitment patterns. We introduce a network-driven labeling method to construct large-scale ground truth for trafficking-at-risk job advertisements. Our results reveal significant linguistic differences between safe and risky advertisements and demonstrate that language models and embedding representations behave distinctly across these linguistic spaces. Building on these insights, we propose a multi-model ensemble classifier to improve the detection of trafficking-at-risk job ads. Finally, we analyze the geographic, gender, industry, and contact-method preferences of trafficking recruiters, revealing systematic patterns in recruitment strategies. 
\end{abstract}


\section{Introduction}

Human trafficking remains a persistent global challenge, one that has been significantly exacerbated in the digital age. Social media platforms and online marketplaces increasingly facilitate connections between traffickers and potential victims by lowering barriers to contact and enabling anonymity \cite{fraser_analysis_2016}. Platforms originally designed for communication, job searching, and social networking are frequently repurposed for illicit recruitment and exploitation \cite{sarkar_use_2015, latonero_human_2011}, placing individuals at risk of harm \cite{allen_role_2019}. Community-oriented websites such as \textit{chineseinla.com} exemplify this dual-use dynamic: while serving as critical support infrastructures for diasporic communities, they may also inadvertently enable trafficking-related recruitment activities.

Recruitment constitutes a pivotal stage in the human trafficking pipeline, marking the initial point at which traffickers identify, contact, and lure potential victims. Recent reports from both the United Nations Office on Drugs and Crime (UNODC) and the U.S. Department of State emphasize the growing prevalence of online recruitment between 2012 and 2022, with job boards and dating platforms serving as primary vectors \cite{united_nations_office_on_drugs_and_crime_unodc_global_2024, office_to_monitor_and_combat_trafficking_in_persons_trafficking_2023}. Although academic research has begun to document how traffickers exploit online platforms for recruitment \cite{moyo_investigating_2025}, comparatively little work has examined scalable computational approaches for regulating or moderating these recruitment messages at their point of origin \cite{greiman_emergence_2013, latonero_rise_2012, ibanez_detection_2014}.

Existing technical and policy interventions predominantly emphasize victim-centered protections, such as background-check mechanisms on dating platforms or automated systems for detecting suspicious escort advertisements \cite{whitney_dont_2018, tong_combating_2017, zhu_identification_2019, lee_infoshield_2021, tobey_interpretable_2024, khan_effects_2023, palmquist_17_2023}. While valuable, these approaches often operate downstream—after initial contact has already occurred—and thus provide limited leverage for preventing exploitation at earlier stages.

In contrast, proactively identifying traffickers who masquerade as legitimate recruiters remains underexplored. This paper addresses this gap by focusing on the recruitment stage of trafficking and leveraging network analysis and natural language processing to trace traffickers’ online footprints across job advertisement ecosystems. By shifting attention from victim detection to recruiter identification, we aim to enable earlier intervention and more effective platform governance.

\subsection{Contributions}
This paper makes the following contributions:
\begin{enumerate}
    \item We expand and annotate a large-scale dataset of 85{,}841 unique job advertisements from non-mainstream online job boards.
    \item We propose a network-driven labeling framework to identify potential human traffickers at the early recruitment stage.
    \item We demonstrate substantial linguistic differences between safe and risky job advertisements.
    \item We provide empirical evidence that language models and embedding methods exhibit systematically different behaviors across safe and trafficking-associated linguistic spaces.
    \item We introduce a classification model that improves the identification of risky job advertisements.
    \item We analyze structural attributes of risky recruitment, including preferred locations, gender specifications, industries, and contact methods used by trafficking-at-risk recruiters.
    \item we open source our data and model to support more research combating human trafficking.
\end{enumerate}

\section{Related Work}

Human trafficking is a complex and adaptive crime that increasingly operates through online platforms, where traffickers exploit the affordances of social media, messaging applications, and online marketplaces to identify, contact, and recruit potential victims \cite{united_nations_report_2003, united_nations_protocol_2000, melton_what_2020}. Prior research has documented how platforms originally designed for social interaction, employment seeking, or community building are repurposed for trafficking-related activities, including deceptive job recruitment, grooming, and coercive communication \cite{latonero_human_2011, sarkar_use_2015, fraser_analysis_2016}. In response, much of the existing scholarship and intervention work has focused on victim-centered detection and support, often emphasizing qualitative analyses of survivor narratives, recruitment tactics, and platform governance \cite{greiman_emergence_2013, allen_role_2019}.

Among computational approaches to detecting criminal activity online, language-based methods have played a central role. Early work in computational criminology demonstrated the utility of text mining for identifying illicit behavior in noisy online environments. For example, \citet{siddiqui_criminal_2019} outlined a pipeline combining preprocessing, classification, clustering, and topic modeling to detect suspicious discourse on social media. Similarly, \citet{bache_language_2010} applied probabilistic language models to crime narratives, showing that word distributions in police reports can predict offender characteristics such as age and gender. Together, these studies highlight how linguistic signals can function as interpretable proxies for criminal style and behavioral patterns.

In the context of human trafficking, language models have primarily been applied to identify victims or downstream exploitation activities. Existing work has trained deep multimodal neural networks on escort advertisements to detect coordinated trafficking networks and sex trafficking operations \cite{tong_combating_2017, lee_infoshield_2021, zhu_identification_2019, li_detection_2018}. Complementary efforts analyze the use of emojis and coded language in sex advertisements \cite{whitney_dont_2018}, or develop interpretable models from illicit massage business reviews \cite{tobey_interpretable_2024}. While effective at identifying trafficking-related activity after recruitment has occurred, these approaches largely operate downstream, when victims may have already been exposed to harm.

More broadly, existing interventions tend to fall into two categories: profiling offenders based on social media traits, or safeguarding potential victims through platform-level interventions. Relatively little attention has been paid to job advertisements on non-mainstream job boards, which often lack formal moderation and remain susceptible to exploitation. In line with calls from UNODC, many preventive efforts rely heavily on user awareness and self-reporting mechanisms. Mainstream platforms such as LinkedIn and Indeed encourage reporting of suspicious job postings \cite{khan_effects_2023, palmquist_17_2023}, while dating platforms such as Tinder and Match have introduced background checks when offline meetings are proposed.

On the regulation and moderation front, however, systematic computational solutions remain limited—particularly for non-mainstream platforms. Prior work has shown that job postings on these sites frequently rely on phone-based contact methods \cite{zhou_tracing_2024}, with industries such as childcare, restaurants, and massage services exhibiting significantly higher reliance on phone contact compared to sectors such as information technology \cite{zhou_tracing_2024}. This informality not only obscures accountability but also provides a novel signal for tracing trafficking-related recruitment networks.

Recent qualitative and mixed-methods research further underscores the relevance of linguistic cues in recruitment-stage trafficking. Through ethnographic analysis across 14 social media platforms and interviews with stakeholders, \citet{moyo_investigating_2025} demonstrate how specific linguistic features can signal risk in job advertisements. and argue for the development of computational approaches to identify traffickers through language patterns.

Together, existing studies reveal a critical gap in preventive, recruiter-focused detection during the recruitment phase of trafficking, while automated moderation tools for this purpose remain scarce. Risk signals embedded in linguistic features and contact information offer a promising opportunity to classify whether a job advertisement is at risk of human trafficking. Building on this insight, our study takes an initial step toward addressing this gap by developing a language-model–based approach to identify high-risk job advertisements before exploitation occurs. Beyond enabling early intervention, our model uncovers hidden linguistic and structural patterns in deceptive recruitment practices, shedding light on how traffickers mask illicit intentions under the rhetoric of legitimate labor.
\begin{figure*}[t]
    \centering
    \includegraphics[width=\textwidth]{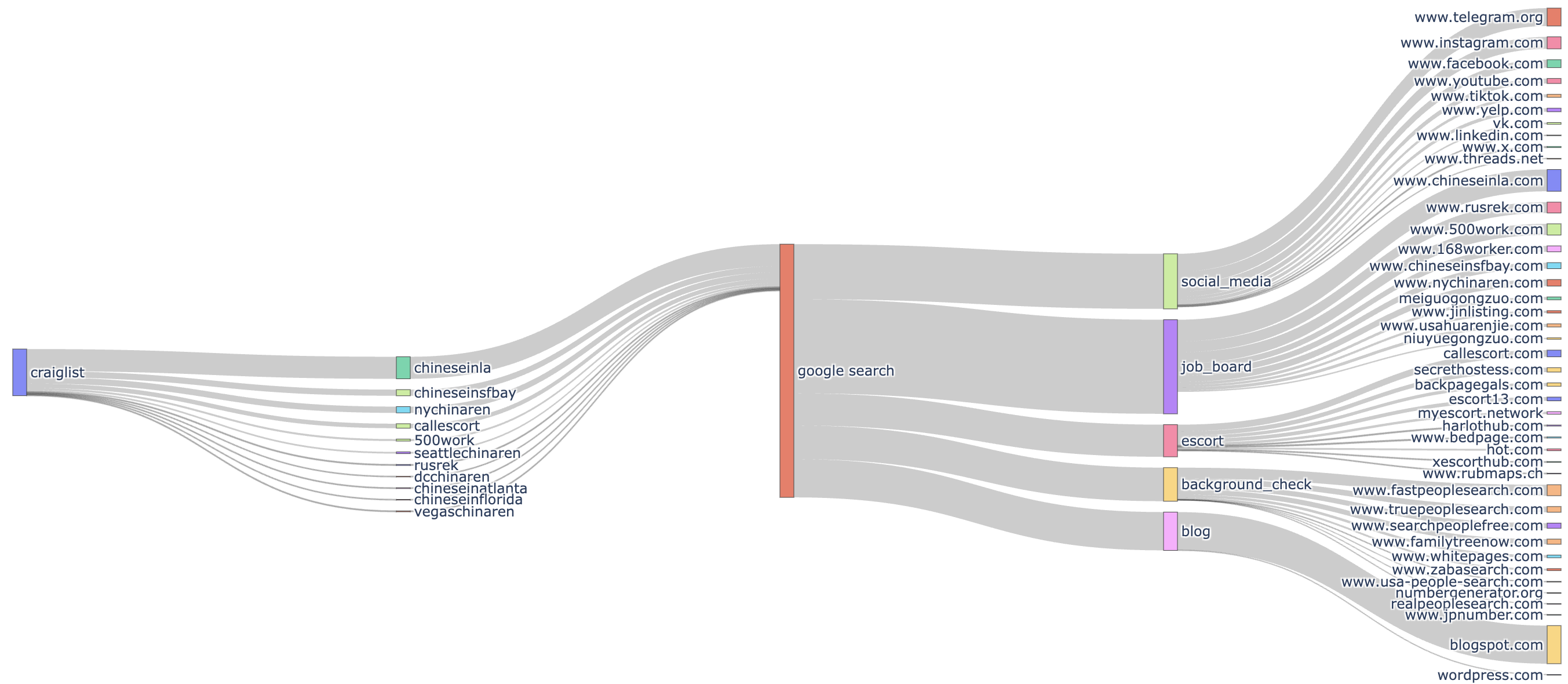}
    \caption{Process of our snowball sampling}
    \label{fig:wide_figure}
\end{figure*}

Building on this motivation, we address the following research questions:

\noindent \textbf{RQ1.} Are the linguistic patterns of risky job advertisements significantly different from those of safe job advertisements?\\
\textbf{RQ2.} How accurately can risky job advertisements be predicted based on linguistic features alone?\\

Assuming effective classification, we further examine trafficking-related recruitment behaviors by addressing the following questions:

\noindent \textbf{RQ3.} Which locations are more frequently associated with risky job advertisements?\\
\textbf{RQ4.} Which industries are disproportionately represented among risky job advertisements?\\
\textbf{RQ5.} Which gender preferences are favored by risky job advertisements?\\
\textbf{RQ6.} Which contact methods are preferred by risky job advertisements?




\section{Method}
To construct a reliable training and evaluation dataset, we adopt a labeling strategy grounded in established practices within law enforcement and anti-trafficking investigations. Prior work and institutional reports have shown instance that traffickers and intermediaries  reuse contact information, particularly phone numbers, across multiple platforms to coordinate recruitment and advertising activities \cite{keegan_using_2019, varadarajan_imbwatch_2025}. This cross-platform linkage is not only a well-documented behavioral pattern but also a standard investigative heuristic used by law enforcement agencies to identify potential trafficking operations \cite{united_nations_office_on_drugs_and_crime_unodc_global_2024,federal_bureau_of_investigation_fbi_human_2020,office_to_monitor_and_combat_trafficking_in_persons_trafficking_2023}. By aligning our labeling approach with these real-world investigative practices, we ensure that our operationalization of trafficking risk reflects domain-relevant signals that extend beyond purely textual features and captures structurally meaningful patterns of coordination across online ecosystems.
\subsection{Data collection}
We collect our samplesOur data collection employed a snowball sampling strategy to source web data across 17 distinct domains. We began by identifying 96 job advertisements from Craigslist. However, due to Craigslist’s strict anti-scraping policies, we did not crawl its listings directly. Instead, we located 11 external domains that had reposted the same job advertisements originally found on Craigslist. These sites served as our seeding domains. Among them were 10 job-board websites in Chinese and Russian advertising employment opportunities in the United States, and one escort website that shared phone contact information with a Craigslist job ad—an early indicator of potential risk linkage.

For each seeding domain, we collected all available job or escort advertisements, extracting any phone numbers listed as contact methods. These phone numbers then formed the core of our cross-platform search. Using the SerpAPI interface to Google Search, we queried each number to locate webpages where it co-occurred, scraping those pages when still publicly available. The specific process is shown in figure ~\ref{fig:wide_figure}. In total, this process yielded 187,213 posts from the 11 seeding domains, along with 126,718 domains and 1,739,090 posts retrieved from Google search results, respectively, across two rounds of expansion (See Appendix Table 3 for detailed data summary).

\subsection{Data Cleaning and Filtering}

To ensure data integrity, we implemented a two-step filtering pipeline based on exact match validation and domain frequency ranking. First, we conduct an exact match filtering. Each Google search result returns a short snippet—a substring of the webpage that the search engine considers a match to the query. We verified whether our queried phone number was indeed a substring within this snippet. Results failing this condition were excluded, as they did not constitute true matches.

Next, we apply domain frequency filtering strategy. We ranked all domains by the number of unique webpages retrieved from each. Domains contributing fewer than five valid posts were excluded to remove noise from rarely indexed or inactive sites.

After cleaning, we categorized the remaining domains using a keyword-based snowballing approach. We began with seed keyword lists derived from common knowledge and observed linguistic patterns in our seeding domains. For instance, keywords such as “sex” and “escort” were characteristic of escort-related sites, while “work,” “ren”, and “chinesein” frequently appeared in job-board domains. We also prelisted known social media platform names as identifiers for that category.

We then grouped domains based on keyword overlaps and manually inspected uncategorized domains, iteratively expanding our keyword lexicon until all remaining sites were classified. This process resulted in a final taxonomy of 1,755 domains, grouped into five major categories (as shown in Figure \ref{fig:wide_figure}).

\subsection{Labeling}
To construct a training and testing dataset for detecting job advertisements at risk of human trafficking, we developed a network-based labeling framework based on several existing research. Existing studies have shown that  traffickers and intermediaries often reuse contact information, particularly phone numbers, across multiple platforms to recruit victims and advertise services, creating observable cross-platform linkages between seemingly distinct domains (e.g., job boards and escort websites); law enforcement routinely leverages this pattern to discover potential trafficking cues \cite{united_nations_office_on_drugs_and_crime_unodc_global_2024}. While not all trafficker reuse their contact information for recruitment and advertisement, obtaining a portion of the observable samples allow us to train language models to identify linguistic patterns and potentially predict those more hidden ones. Building on this insight, we operationalize trafficking risk as the cross-domain reuse of phone numbers between labor-oriented and sex-oriented platforms.

Specifically, we constructed a phone-number co-occurrence network, linking each advertisement to the phone numbers it listed and the domains on which those numbers appeared. Using previously established domain-level risk labels, we assigned post-level labels as follows:

Risky: if a phone number associated with a job advertisement also appeared on escort websites;

Safe: otherwise. 

This approach captures a structural signal of potential coordination between recruitment and commercial sex advertising, rather than relying solely on textual indicators, which may be intentionally obfuscated.

We acknowledge that this proxy does not directly observe trafficking outcomes, but instead identifies risk exposure based on cross-platform behavioral patterns. To assess the validity of this labeling strategy, we conducted a manual audit of a stratified random sample of labeled advertisements (N = 420). Human coders evaluated whether the content exhibited indicators commonly associated with trafficking-related recruitment listed by existing studies(e.g., vague job descriptions, high compensation promises, off-platform contact requests) \cite{volodko_spotting_2020}. The results show that 95.6\% (239 out of 250) of ads labeled as risky contained at least one trafficking-related indicator, compared to 5.3\% (9 out of 170) in the safe group, providing preliminary support for the construct validity of our proxy. See Examples of safe and risky job ads in appendix. 

To ensure sample uniqueness, we removed duplicate advertisements and retained only one instance per post. The initial dataset contained 1,605 risky posts, resulting in substantial class imbalance. To address this, we expanded the risky class by incorporating additional webpages collected via Google search that shared risk-labeled phone numbers, increasing the number of risky samples to 2,816. This augmentation leverages the same cross-platform linkage assumption and preserves consistency in the labeling logic. Table \ref{tab:domains} presents the distribution of samples across categories.

\subsection{Sampling}

We experimented with two sampling strategies to mitigate class imbalance during model training, both with randomseed = 42:

\textbf{Moderate Oversampling (80/20)}
We subsampled the safe class so that risky samples constituted 20\% of the total training data (up from 3.3\%). This approach preserved a large overall dataset while increasing the relative influence of risky samples.

\textbf{Balanced Sampling (50/50)}
We further downsampled the safe posts to achieve an equal ratio between risky and safe samples, resulting in a balanced training dataset of 5,632 total posts. This setup allowed us to evaluate model sensitivity and stability under perfectly balanced conditions.

We compare model performance under these two strategies in the Results section to assess the trade-off between dataset balance and overall sample size to answer \textbf{RQ2}.

\subsection{Modeling and Evaluation}

Our modeling pipeline is designed to systematically address RQ1 and RQ2, with a focus on evaluating how different language models and embedding representations identify risky job advertisements across multilingual contexts. 

Although explicit linguistic features listed in \citet{moyo_investigating_2025} are more interpretable with human nature, it would be difficult to extract those human summarized features in job ads. In addition, we could be missing some of the other less explicit linguistic features if we stick to a set of fixed ones. Therefore, we use a machine learning approach to embed the context of the job ads into 4096 features, and we interpret and analyze the meaning of those features after our model successfully classified all safe and high-risk job ads. To ensure phone numbers which we used for labeling do not spoil the learning process, we removed all the phone numbers in the job ads during the training stage. 

\subsubsection{Embedding Comparison}

To assess embedding performance for multilingual job advertisements, we evaluate two state-of-the-art multilingual embedding models: \textbf{BGE-M3} and \textbf{Qwen3}. Both models are designed for cross-lingual semantic representation and have demonstrated strong performance across multiple languages, making them well suited for our dataset spanning English, Chinese, and Russian posts.

For consistency, embeddings were generated using a batch size of \textbf{2} and a maximum sequence length of \textbf{4096 tokens}. All embeddings were conducted on a single \textbf{NVIDIA L40S GPU} using CUDA-enabled inference. Unless otherwise specified, we use the default model configurations provided by the respective implementations.

We first visualize the latent representations produced by each embedding model using Principal Component Analysis (PCA) to answer \textbf{RQ1}. This visualization allows us to examine whether risky and safe job advertisements form distinguishable clusters in the embedding space—an indicator of linguistic separability between the two classes. Clearer cluster boundaries suggest that the embedding model captures meaningful semantic and stylistic differences relevant to trafficking risk.

To quantitatively assess embedding effectiveness, we compare downstream classification performance using confusion matrices, ROC-AUC, and PR-AUC. These metrics allow us to evaluate not only overall classification accuracy but also each model’s ability to detect low-prevalence risky content under class imbalance conditions that are characteristic of illicit recruitment data.

\subsubsection{Classification Models}

To evaluate the predictive performance of existing modeling approaches and the impact of resampling strategies, we test five classes of models that span a range of interpretability, complexity, and data requirements:

\textbf{Few-shot Large Language Models (LLMs)} We evaluate zero-shot and few-shot performance using a pre-trained LLM without additional fine-tuning. The few-shot setup includes k = \{1,2,5\} in-context examples per class, sampled from the training set. Prompts incorporate wording derived from UN trafficking language guidelines to reflect domain-relevant indicators. All LLM outputs are generated with temperature = 0 to ensure deterministic responses. Full prompt templates and examples are provided in the Appendix.

\textbf{Logistic Regression and XGBoost} Logistic Regression serves as an interpretable baseline to assess linear separability in the embedding space. We use L2 regularization with regularization strength $C=1.0$, optimized via validation.

We extend this baseline with XGBoost, a tree-based ensemble model capable of capturing non-linear decision boundaries. Key hyperparameters include: \textbf{ 1) number of trees = \{200, 400, 600\}, 2) max depth = \{4, 6, 8\}, 3) learning rate = \{0.05, 0.1\}, 4) subsample = \{0.8, 1\}, 5) colsample\_bytree = \{0.8, 1\}}. Hyperparameters are selected via grid search on the validation set.

\textbf{Feedforward Neural Networks (FFNN)} We implement a feedforward neural network to model higher-order non-linear relationships in the embedding space. The network consists of:
\textbf{ 1) two hidden layers (sizes: 256, 128), 2) ReLU activation, 3) dropout rate = 0.2, 4) output layer with sigmoid activation}. Models are trained using the Adam optimizer (learning rate = 1e-3, batch size = 64) for up to 20 epochs, with early stopping based on validation loss.

\textbf{DistilBERT} We fine-tune a pre-trained DistilBERT model for sequence classification. The model is trained using: \textbf{ 1) learning rate = 2e-5, 2) batch size = 16, 3) epochs = 10, 4) Manual BGE and Qwen embedding}. We use the HuggingFace Transformers implementation with default weight initialization and apply early stopping based on validation performance.

All models are implemented in Python using scikit-learn, XGBoost, PyTorch, and HuggingFace Transformers. Random seeds are fixed (seed = 42) to ensure reproducibility.

Each model is trained and evaluated using $5$-fold cross-validation to ensure stability and generalizability. Performance is assessed using confusion matrices, ROC-AUC, and PR-AUC across different resampling strategies (20\% vs.\ 50\% risk balance). These metrics allow us to compare model sensitivity to risky advertisements as well as robustness under varying class distributions.

The aggregated evaluation enables us to determine: (1) whether zero-shot LLMs can effectively identify risky advertisements; (2) which embedding model yields the strongest multilingual representations; and (3) which combination of classifier and sampling strategy achieves the best trade-off among accuracy, recall, and fairness across languages.

We further evaluate each sampling--embedding--model combination with 5-fold cross validation using the F1 score and ROC-AUC for both the safe and risky classes. We report average scores for each model. Based on these results, we design a multi-model ensemble classifier, referred to as the \textit{Trafficker Classifier}, by integrating predictions from all models. Specifically, each model casts a vote on the predicted label, and the final label is assigned based on a majority vote exceeding $50\%$. We evaluate the ensemble using the same F1 and ROC-AUC metrics to assess performance gains over individual models.

\subsection{Characterizing Trafficker-at-Risk Job Advertisements}

We apply the final classifier to distinguish risky job advertisements from safe ones. Using structured attributes extracted from each post—including domain location, job location, phone number location, preferred gender, preferred contact method, and job industry—we compare attribute distributions between safe and risky advertisements to address RQ3–RQ5.

Domain location is inferred from domain names using a keyword-based classifier that matches strings containing city names and their corresponding states. For example, domains such as “chineseinla” and “chineseinsfbay” are labeled as California, while domains without identifiable geographic markers (e.g., “500work”, “rusrek”) are labeled as unspecified.

Job location and job industry are extracted from post content using an iterative keyword snowballing approach. We first curate an initial list of location- and industry-related keywords, apply it to the corpus, and manually inspect several hundred labeled cases. Based on this inspection, we iteratively refine the keyword lists. This process is conducted and validated by researchers fluent in English, Chinese, and Russian to ensure accurate coverage across all languages in the dataset.

Phone number location is identified using an open-source telephone area-code dataset published by ArcGIS Data and Maps.

Preferred gender is classified using rule-based logic: posts mentioning only “male” are labeled as male-preferred; those mentioning only “female” are labeled as female-preferred; posts containing “couple” or “married couple” are labeled as couple-preferred. Posts that do not specify any of these criteria or mention both "male" and "female" are labeled as having no stated gender preference.

For each attribute, we report descriptive statistics in terms of both absolute counts and the proportion of risky advertisements. Absolute volume captures the observed scale of trafficking-related activity within each category, while the proportion of risky advertisements reflects conditional risk—that is, the likelihood that an advertisement is associated with trafficking given specific attributes such as location, preferred gender, contact method, or industry.

\section{Results}
\begin{figure}[h]
    \centering
    \includegraphics[width=0.95\columnwidth]{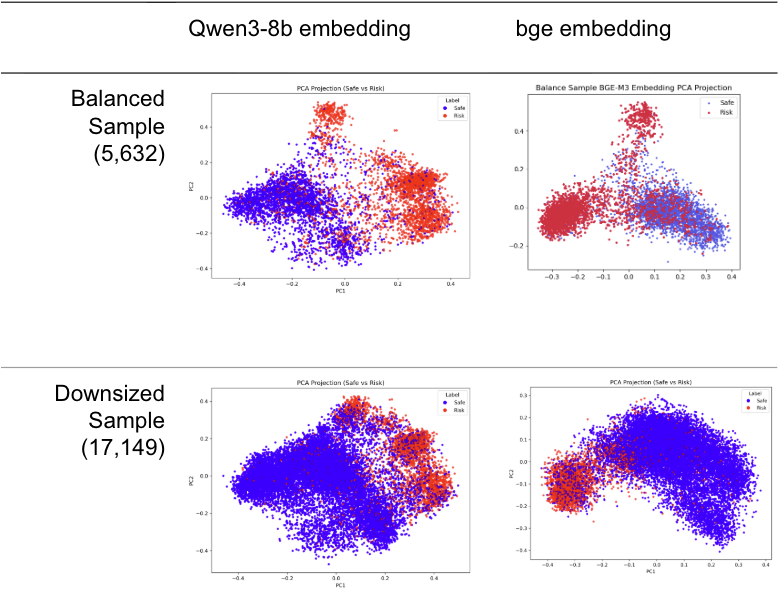}
    \caption{PCA projection visualization for both embeddings and samples}
    \label{fig:QwenPCA}
\end{figure}

\begin{table*}[t]
\centering
\caption{Performance comparison across embeddings, training strategies, and prompt-based methods.}
\label{tab:performance}
\begin{tabular}{llccccc}
\toprule
\textbf{Embedding} & \textbf{Model} & \textbf{Precision} & \textbf{Recall} & \textbf{Accuracy} & \textbf{ROC-AUC} & \textbf{F1} \\
\midrule
\multicolumn{7}{c}{\textit{Balanced Sample (5,632)}} \\
\midrule
Qwen3-8B & Logistic Reg. & 0.88 & 0.87 & 0.87 & 0.93 & 0.87 \\
 & XGBoost & 0.95 & 0.91 & 0.91 & 0.91 & 0.91 \\
 & FFNN & 0.93 & 0.92 & 0.90 & 0.92 & 0.91 \\
 & DistilBERT & 0.93 & 0.88 & 0.88 & 0.88 & 0.88 \\
\midrule
BGE-M3 & Logistic Reg. & 0.93 & 0.88 & 0.88 & 0.88 & 0.88 \\
 & XGBoost & 0.95 & 0.91 & 0.91 & 0.91 & 0.91 \\
 & FFNN & 0.95 & 0.90 & 0.90 & 0.90 & 0.90 \\
 & DistilBERT & 0.94 & 0.89 & 0.89 & 0.89 & 0.89 \\
\midrule
\multicolumn{7}{c}{\textit{Downsized Sample (17,149)}} \\
\midrule
Qwen3-8B & Logistic Reg. & 0.93 & 0.85 & 0.82 & 0.89 & 0.90 \\
 & XGBoost & \textbf{0.96} & \textbf{0.90} & \textbf{0.92} & 0.88 & \textbf{0.95} \\
 & FFNN & 0.96 & 0.83 & 0.82 & 0.83 & 0.94 \\
 & DistilBERT & 0.94 & 0.88 & 0.89 & 0.88 & 0.94 \\
\midrule
BGE-M3 & Logistic Reg. & 0.93 & 0.84 & 0.81 & 0.88 & 0.90 \\
 & XGBoost & 0.96 & 0.89 & 0.91 & 0.87 & 0.94 \\
 & FFNN & 0.95 & 0.79 & 0.80 & 0.77 & 0.93 \\
 & DistilBERT & 0.94 & 0.87 & 0.88 & 0.86 & 0.93 \\
\midrule
\multicolumn{7}{c}{\textit{Ensemble}} \\
\midrule
Both & \textbf{Trafficker Classifier} & \textbf{0.96} & \textbf{0.99} & \textbf{0.94} & \textbf{0.94} & \textbf{0.96} \\
\midrule
\multicolumn{7}{c}{\textit{Zero/Few-Shot LLM Baselines (Accuracy)}} \\
\midrule
\multicolumn{2}{l}{Qwen3-0.6B} & \multicolumn{5}{c}{0.12 (Zero-shot) / 0.81 (Few-shot)} \\
\multicolumn{2}{l}{Qwen3-8B} & \multicolumn{5}{c}{0.27 (Zero-shot) / 0.42 (Few-shot)} \\
\multicolumn{2}{l}{Deepseek v3.2} & \multicolumn{5}{c}{0.81 (Zero-shot) / 0.79 (Few-shot)} \\
\multicolumn{2}{l}{Gemini} & \multicolumn{5}{c}{0.63 (Zero-shot) / 0.63 (Few-shot)} \\
\bottomrule
\end{tabular}
\end{table*}

\begin{figure*}[h]
    \centering
    \includegraphics[width=\textwidth]{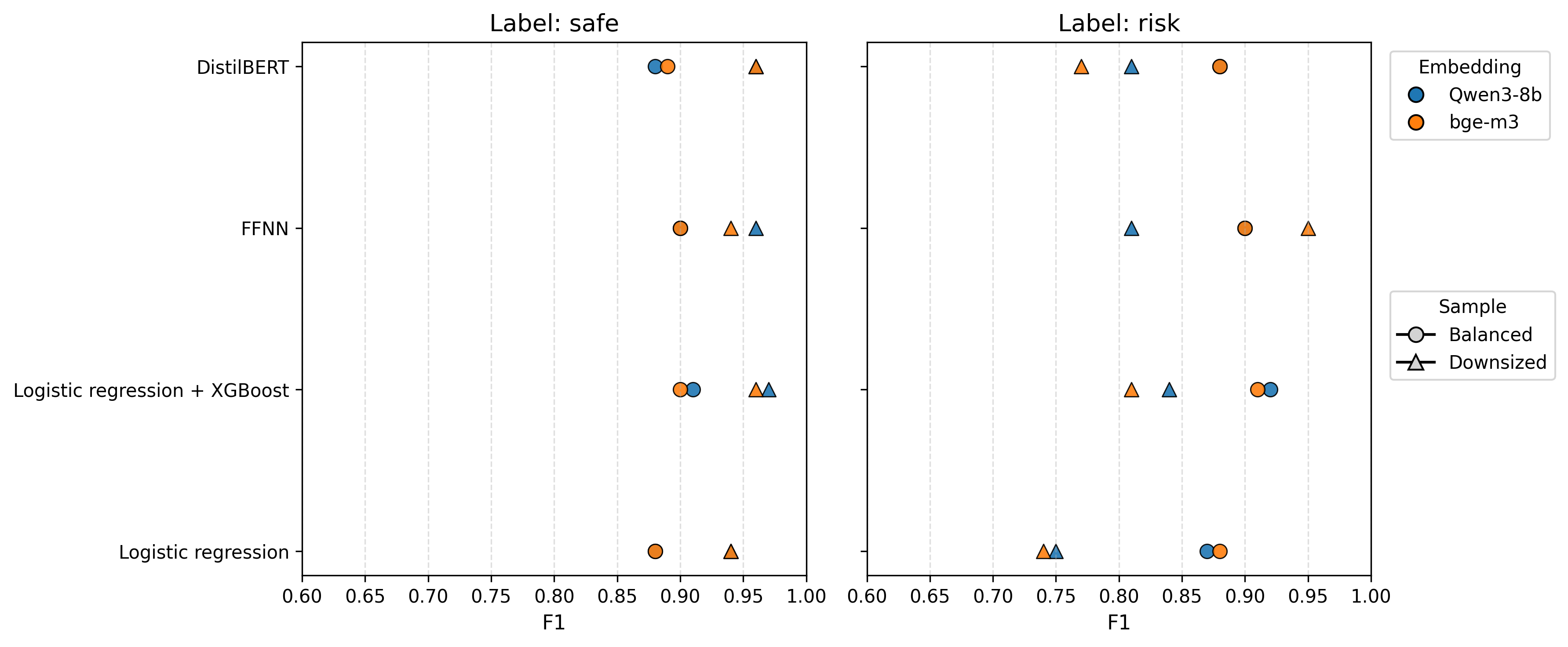}
    \caption{evaluation for each sampling-embedding-model combination on their performance for each label}
    \label{fig:cleve}
\end{figure*}

\subsection{Embedding Separability in Multilingual Space}
Figure~\ref{fig:QwenPCA} presents the two-dimensional PCA projections of embeddings generated by both \textbf{bge-m3} and \textbf{Qwen3} under two sampling strategies (20\% and 50\% risk balance). Across all embedding--sampling combinations, we observe clear and consistent clustering between safe and risky job advertisements. This visual separation indicates that safe and risky posts exhibit systematically different linguistic patterns in the latent semantic space.

Under balanced sampling with \textbf{bge-m3}, a small number of risky posts appear intermixed with safe posts, and under the downsampled settings, both embedding models show limited overlap between safe and risky clusters. These overlaps suggest that a small subset of safe and risky advertisements share similar linguistic features. However, the majority of posts remain well separated into two distinct clusters. The stability of this separation across embedding models and resampling regimes suggests that trafficking-related recruitment language contains robust and learnable semantic signatures that generalize across languages and class distributions.

\subsection{Performance of Individual Classification Models}

We next evaluate the predictive performance of multiple classification models, including Logistic Regression, XGBoost, Feedforward Neural Networks (FFNN), and DistilBERT, and compare them against few-shot large language model (LLM) baselines guided by UN trafficking language prompts.

Across all embedding models and sampling strategies, supervised classifiers consistently outperform few-shot LLM prompting. While the few-shot baselines demonstrate limited sensitivity to subtle trafficking signals, trained classifiers achieve substantially stronger and more stable performance. In general, the supervised models reach precision and recall values near or above 90\% across both safe and risky classes, regardless of embedding choice or sampling ratio. This indicates that once trained on labeled data, both classical machine learning models and neural architectures can reliably distinguish trafficking-related recruitment from legitimate job postings.

Among all model configurations, XGBoost exhibits particularly strong overall performance, achieving an accuracy of 0.95 using the downsampled setting with Qwen embeddings. The remaining model--embedding combinations achieve an average accuracy of 0.92, recall of 0.90, precision of 0.91, and F1 score of 0.90. These results indicate that, while most supervised models perform robustly and consistently, tree-based ensemble methods exhibit a notable advantage under certain sampling and embedding conditions, likely due to their ability to effectively capture non-linear decision boundaries in the embedding space.

\subsection{Model, Sampling, and Label-Specific Trade-offs}

Despite strong overall performance, deeper analysis reveals that no single model--embedding--sampling combination dominates across all evaluation criteria. Instead, distinct combinations exhibit complementary strengths. Shown in figure~\ref{fig:cleve}, downsized bge-m3 distilBERT(F1 = 0.96), downsized Qwen3 FFNN (F1 = 0.96), downsized Qwen3 XGBoost (F1 = 0.97) better labels safe ads while downsized bge-m3 better labels risk ads (F1 = 0.95). That is said some models are better at classifying risky ads, while others are more skilled at labeling safe ads. 

These results suggest that individual classifiers specialize in different regions of the decision space. This heterogeneity in predictive behavior motivates the use of an ensemble strategy that can integrate the strengths of multiple models while compensating for their individual weaknesses.

\subsection{Ensemble Voting and the Trafficker Classifier}

To leverage the complementary strengths of individual classifiers, we construct an ensemble model termed the \textit{Trafficker Classifier}. Under this voting framework, each trained model casts a binary prediction (safe vs.\ risky) for each job advertisement, and the final label is assigned based on a majority vote exceeding $50\%$.

This ensemble approach leads to a substantial performance gain over all individual models. Specifically, the \textit{Trafficker Classifier} achieves a precision of 99\% and a recall of 94\% for the risky class, representing a significant improvement in both false positive control and detection sensitivity. The high precision indicates that nearly all predicted risky advertisements correspond to true trafficking-related content, while the elevated recall demonstrates that the ensemble successfully captures the vast majority of risky cases.

These results confirm that ensemble learning provides a powerful mechanism for stabilizing predictions across multilingual and imbalanced settings, offering a robust and deployable solution for large-scale trafficking risk detection.

\begin{figure*}[h]
\centering

\begin{subfigure}[b]{0.48\textwidth}
  \centering
  \includegraphics[width=\textwidth, height=0.625\textwidth]{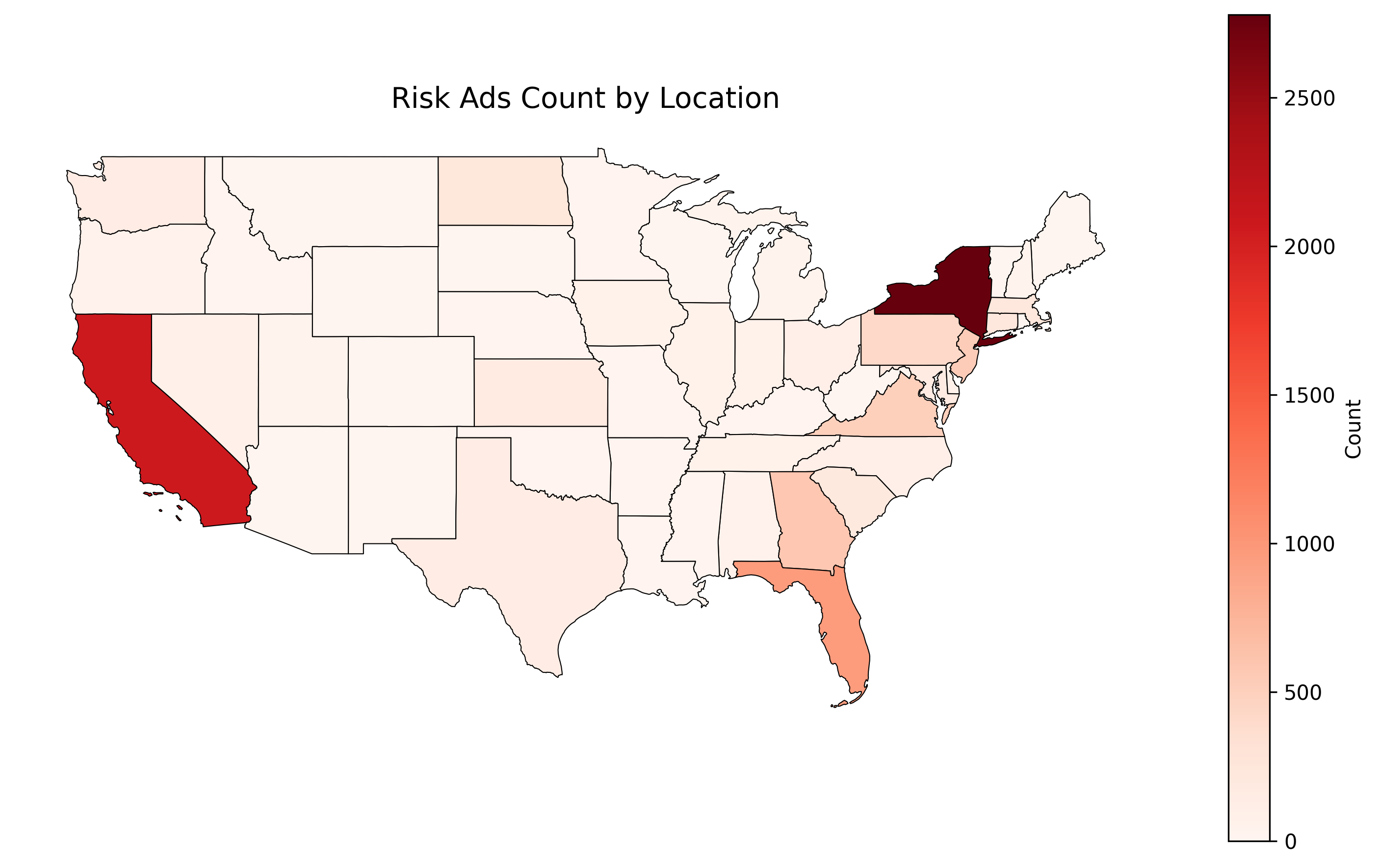}
  \caption{Volume}
  \label{fig:location-volume}
\end{subfigure}
\hfill
\begin{subfigure}[b]{0.48\textwidth}
  \centering
  \includegraphics[width=\textwidth, height=0.625\textwidth]{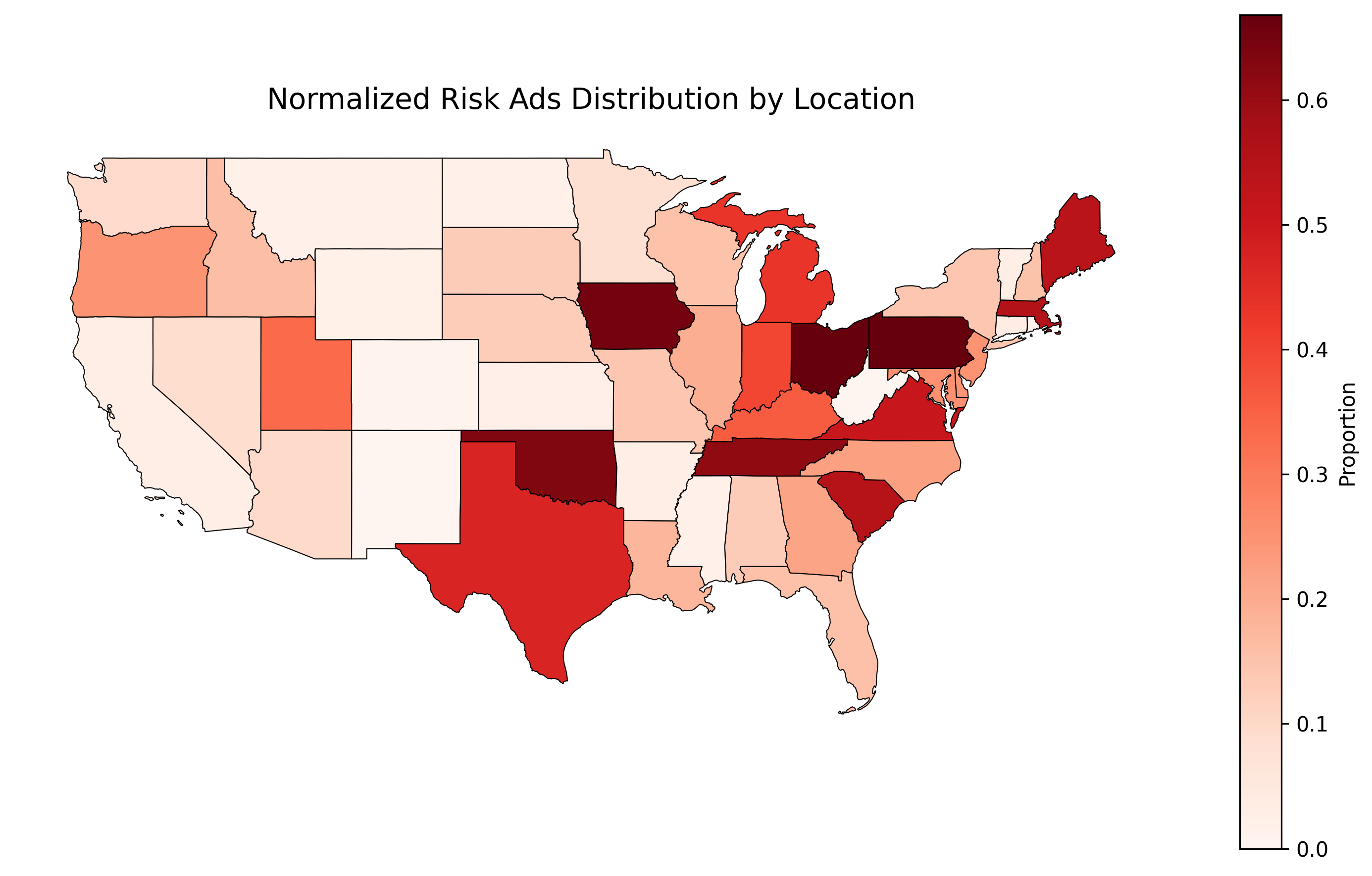}
  \caption{Percentage}
  \label{fig:location-percent}
\end{subfigure}

\caption{Distribution of where ads claim applicants will be working, shown by (left) volume and (right) percentage.}
\label{fig:location-distribution}
\end{figure*}












\begin{table}[h]
\centering
\caption{Location Matching Statistics by Category and Risk Type}
\label{tab:location_match_transposed_percent}

\small
\begin{tabular}{@{}lrr@{}}
\toprule
\textbf{Category} & \textbf{Safe (\%)} & \textbf{Risky (\%)} \\
\midrule
All Match & 41,042 (26.8) & 2,069 (18.8) \\
All Mismatch & 4,053 (2.6) & 190 (1.7) \\
Phone Loc Mismatch & 4,495 (2.9) & 630 (5.7) \\
Domain Loc Mismatch & 1,236 (0.8) & 372 (3.4) \\
Job Loc Mismatch & 22,136 (14.4) & 700 (6.4) \\
Domain Loc Unspec (Match) & 4,941 (3.2) & \textbf{4,080 (37.1)} \\
Domain Loc Unspec (Mismatch) & 3,765 (2.5) & \textbf{1,881 (17.1)} \\
Phone Loc Unknown (Match) & \textbf{37,058 (24.2)} & 532 (4.8) \\
Phone Loc Unknown (Mismatch) & \textbf{27,694 (18.1)} & 218 (2.0) \\
Job Loc Unknown (Match) & 2,654 (1.7) & 134 (1.2) \\
Job Loc Unknown (Mismatch) & 546 (0.4) & 60 (0.5) \\
Not Comparable & 3,801 (2.5) & 141 (1.3) \\
\bottomrule
\end{tabular}
\end{table}

\begin{figure}[h]
\centering
\begin{minipage}[b]{0.48\textwidth}
  \centering
  \includegraphics[width=\textwidth, height=0.625\textwidth]{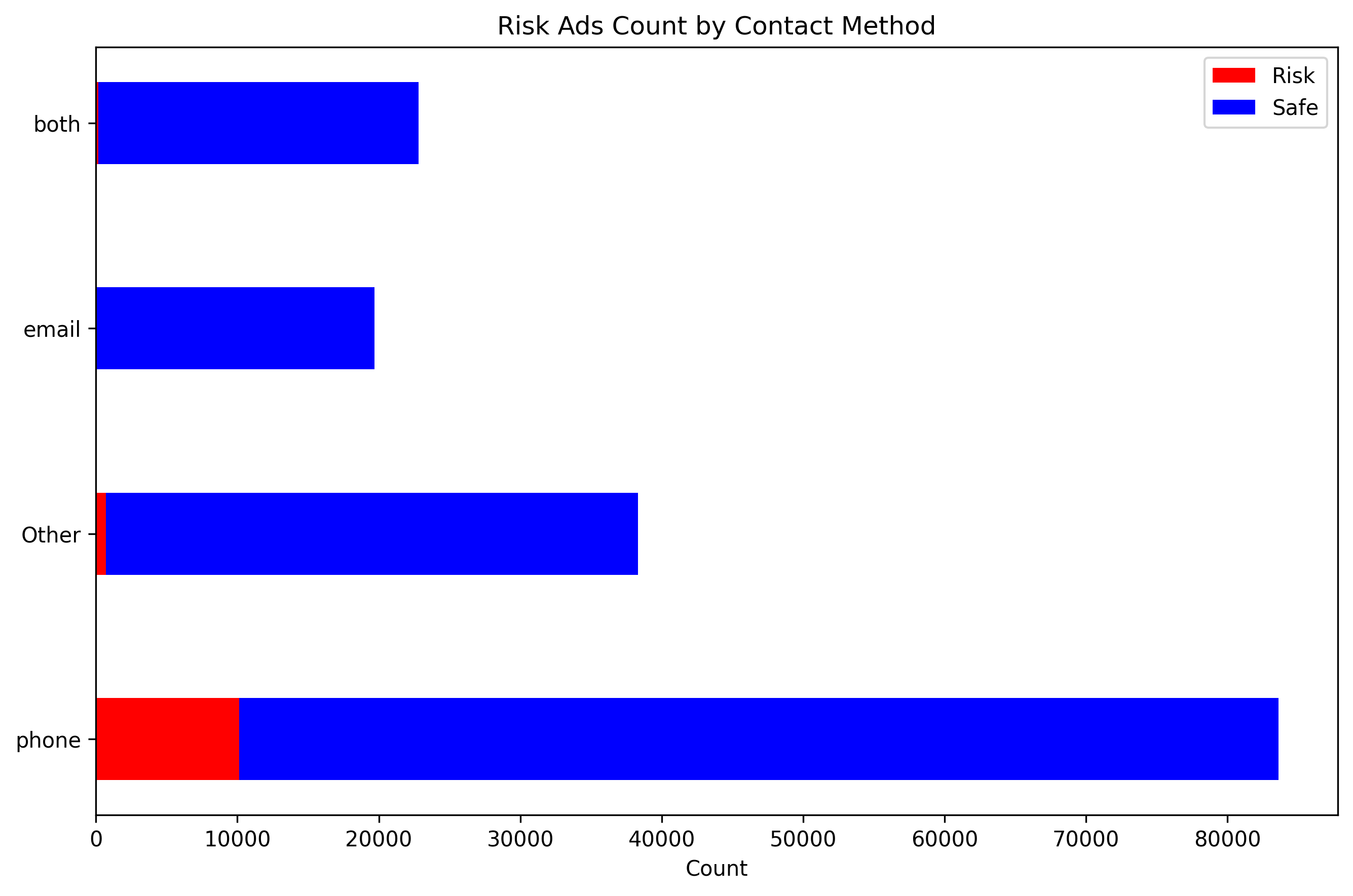}
  \caption{Distribution of preferred contact methods in volume}
  \label{fig:first}
\end{minipage}%
\hfill
\begin{minipage}[b]{0.48\textwidth}
  \centering
  \includegraphics[width=\textwidth, height=0.625\textwidth]{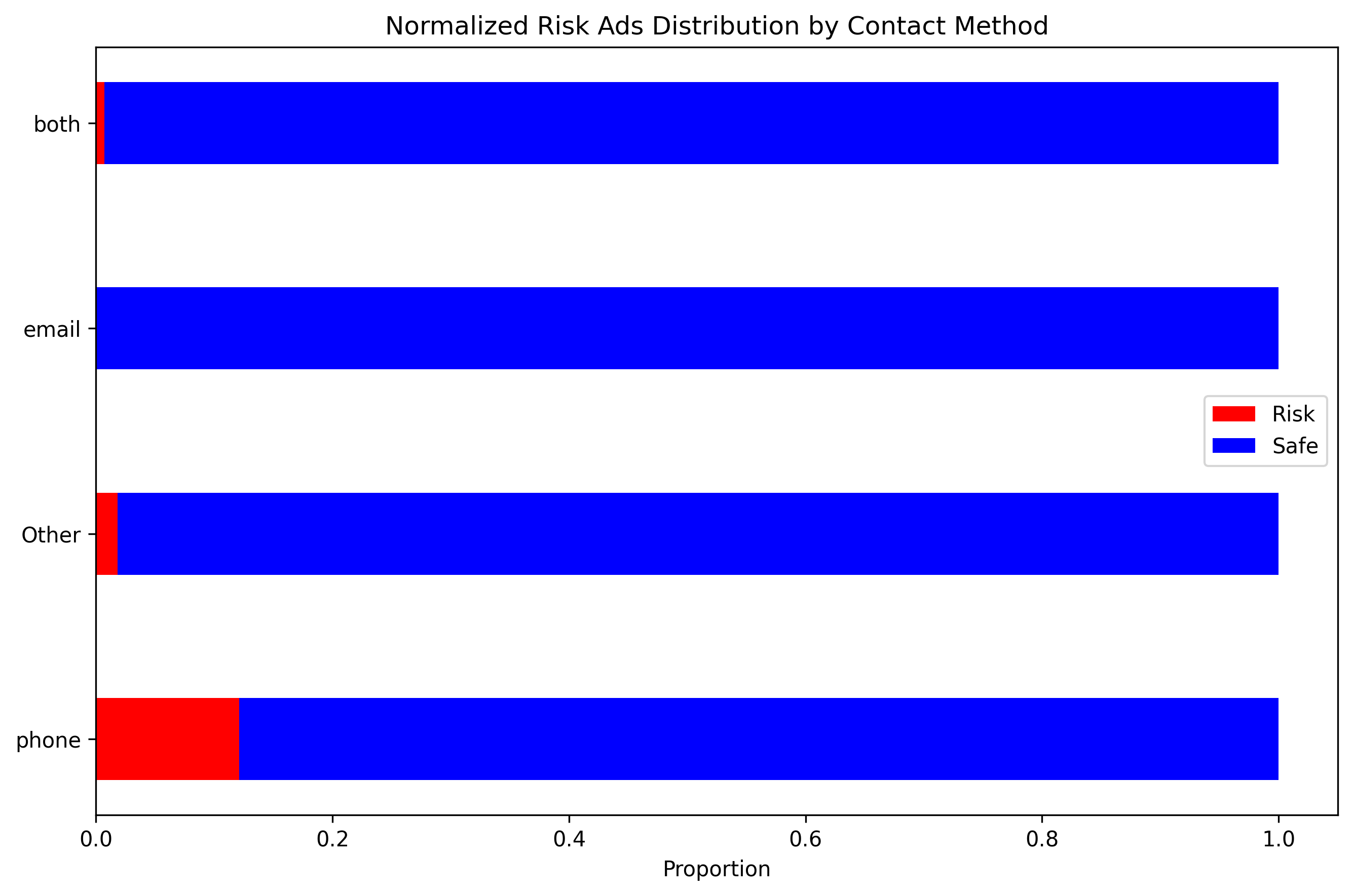}
  \caption{Distribution of preferred contact methods in percentage}
  \label{fig:second}
\end{minipage}
\end{figure}

\begin{figure}[h]
\centering
\begin{minipage}[b]{0.48\textwidth}
  \centering
  \includegraphics[width=\textwidth, height=0.625\textwidth]{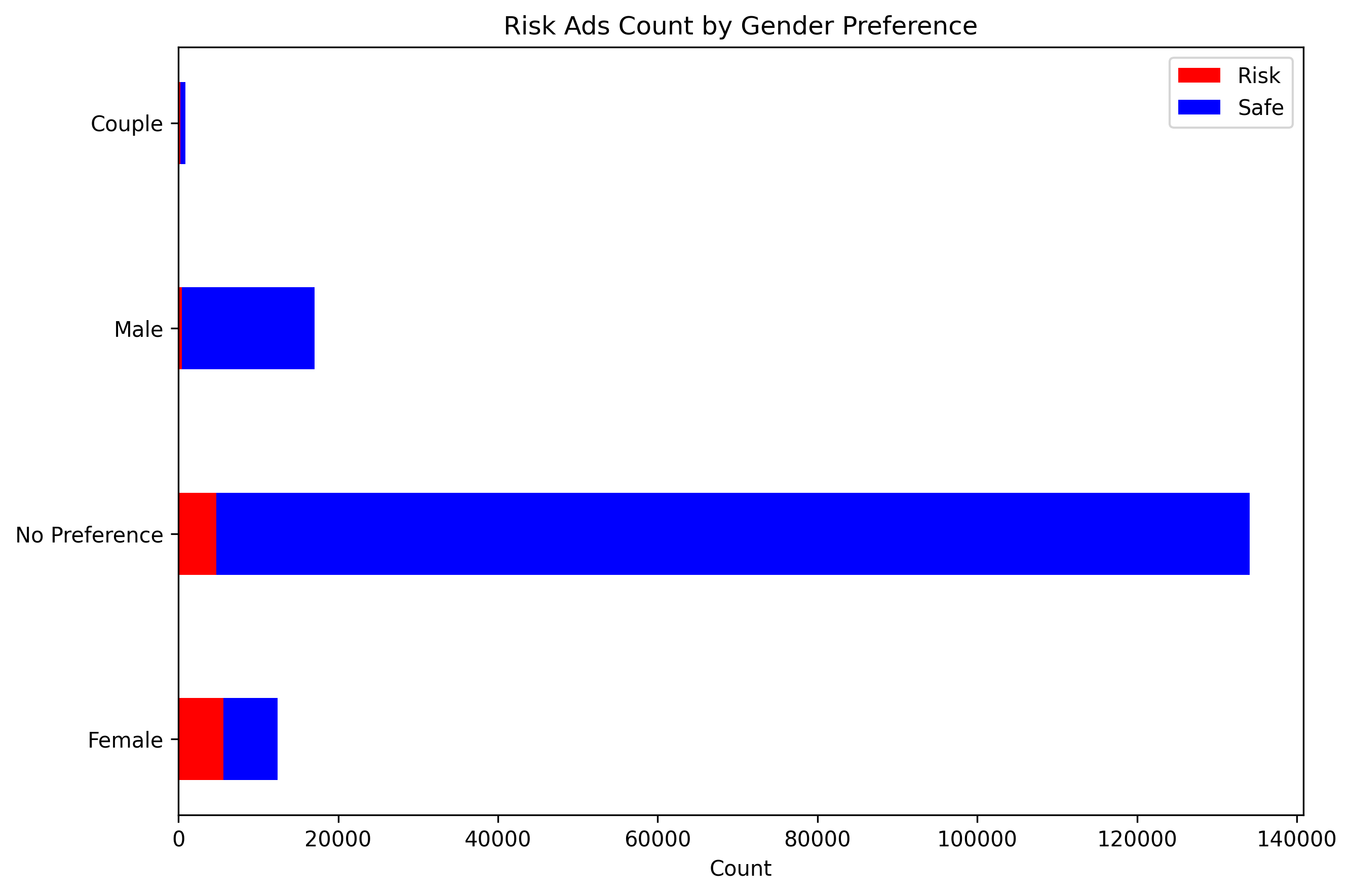}
  \caption{Distribution of gender preference in volume}
  \label{fig:first}
\end{minipage}%
\hfill
\begin{minipage}[b]{0.48\textwidth}
  \centering
  \includegraphics[width=\textwidth, height=0.625\textwidth]{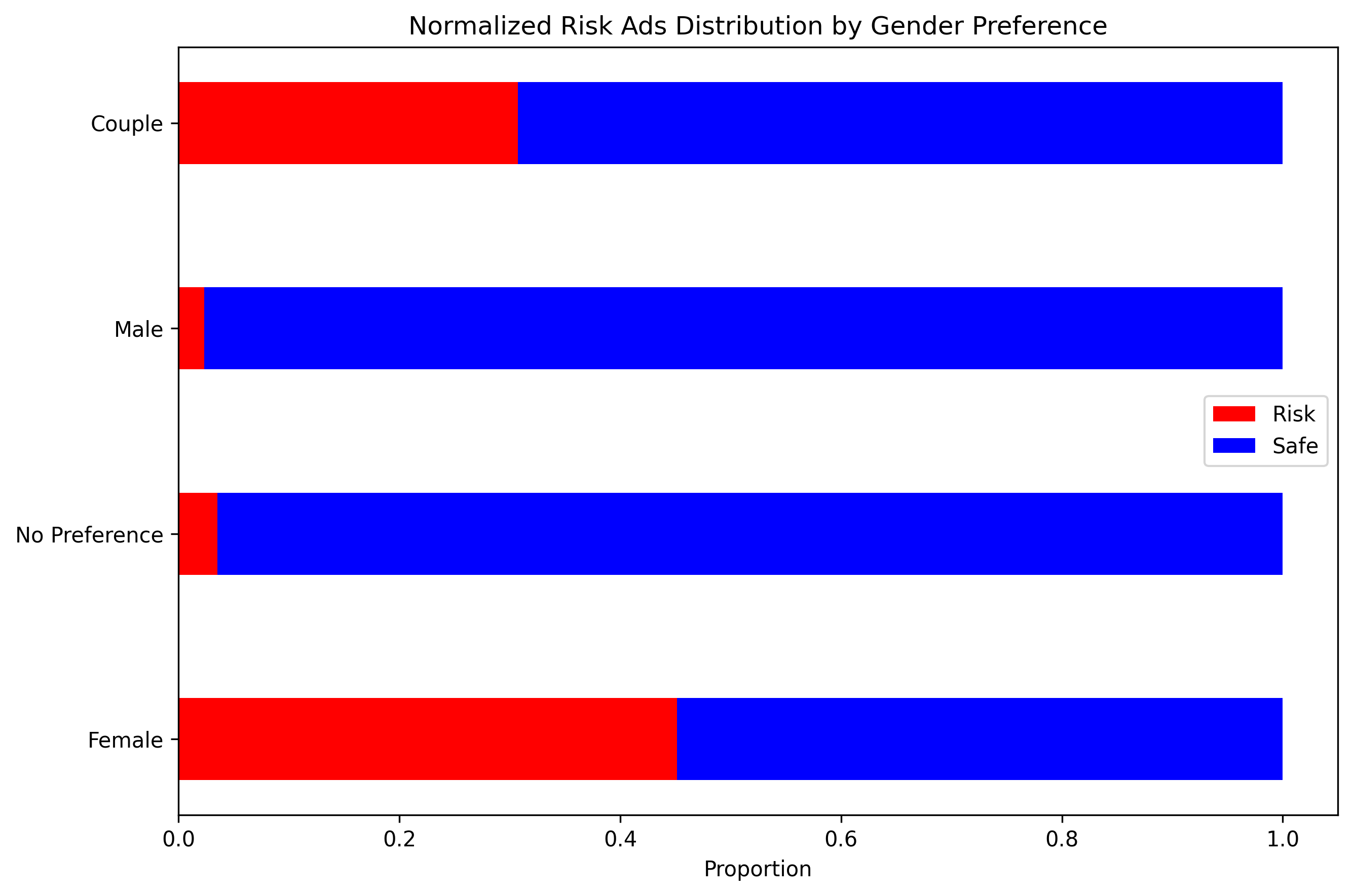}
  \caption{Distribution of gender preference in percentage}
  \label{fig:second}
\end{minipage}
\end{figure}

\begin{figure}[h]
\centering
\begin{minipage}[b]{0.48\textwidth}
  \centering
  \includegraphics[width=\textwidth, height=0.625\textwidth]{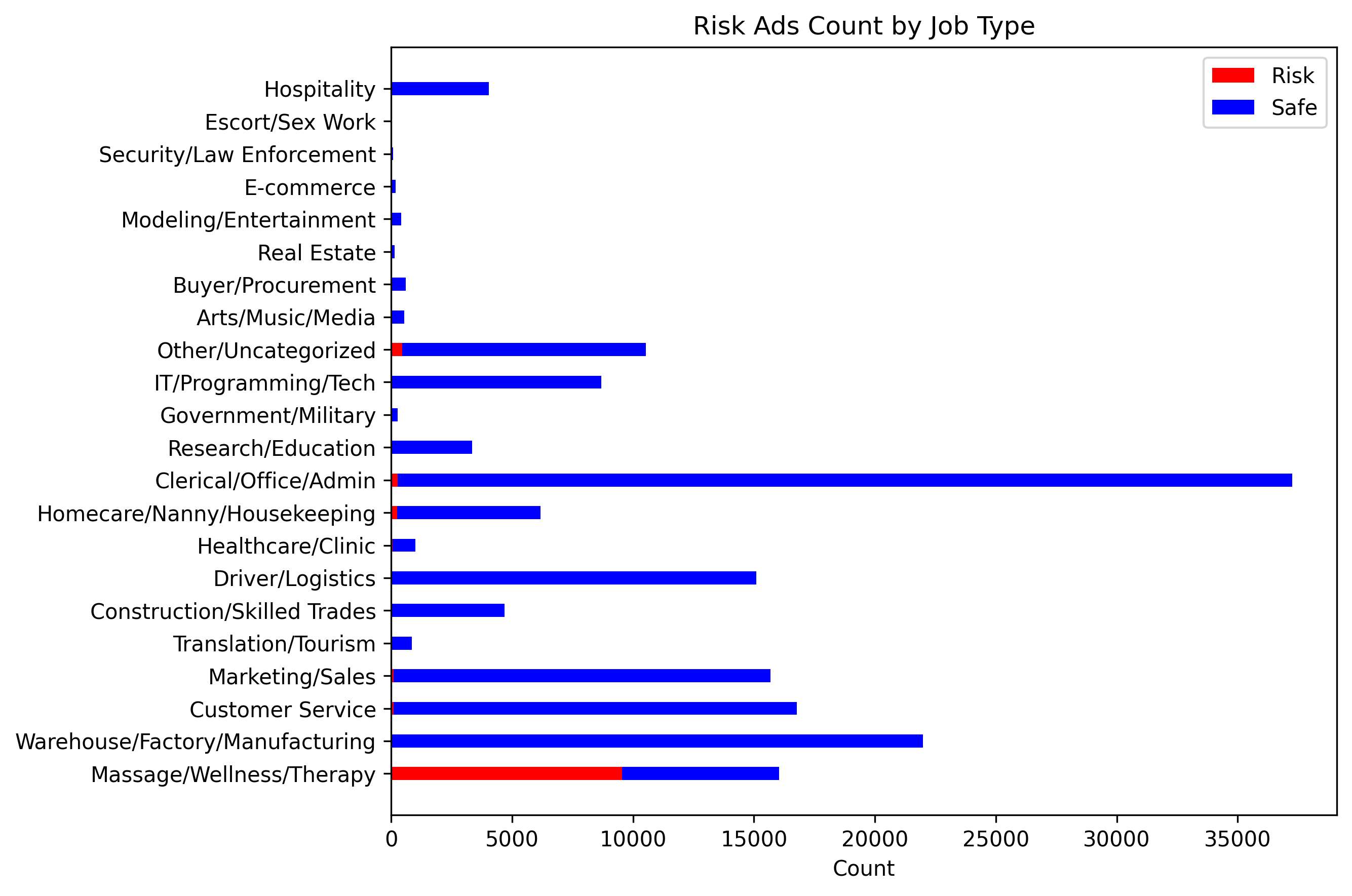}
  \caption{Distribution of industry in volume}
  \label{fig:first}
\end{minipage}%
\hfill
\begin{minipage}[b]{0.48\textwidth}
  \centering
  \includegraphics[width=\textwidth, height=0.625\textwidth]{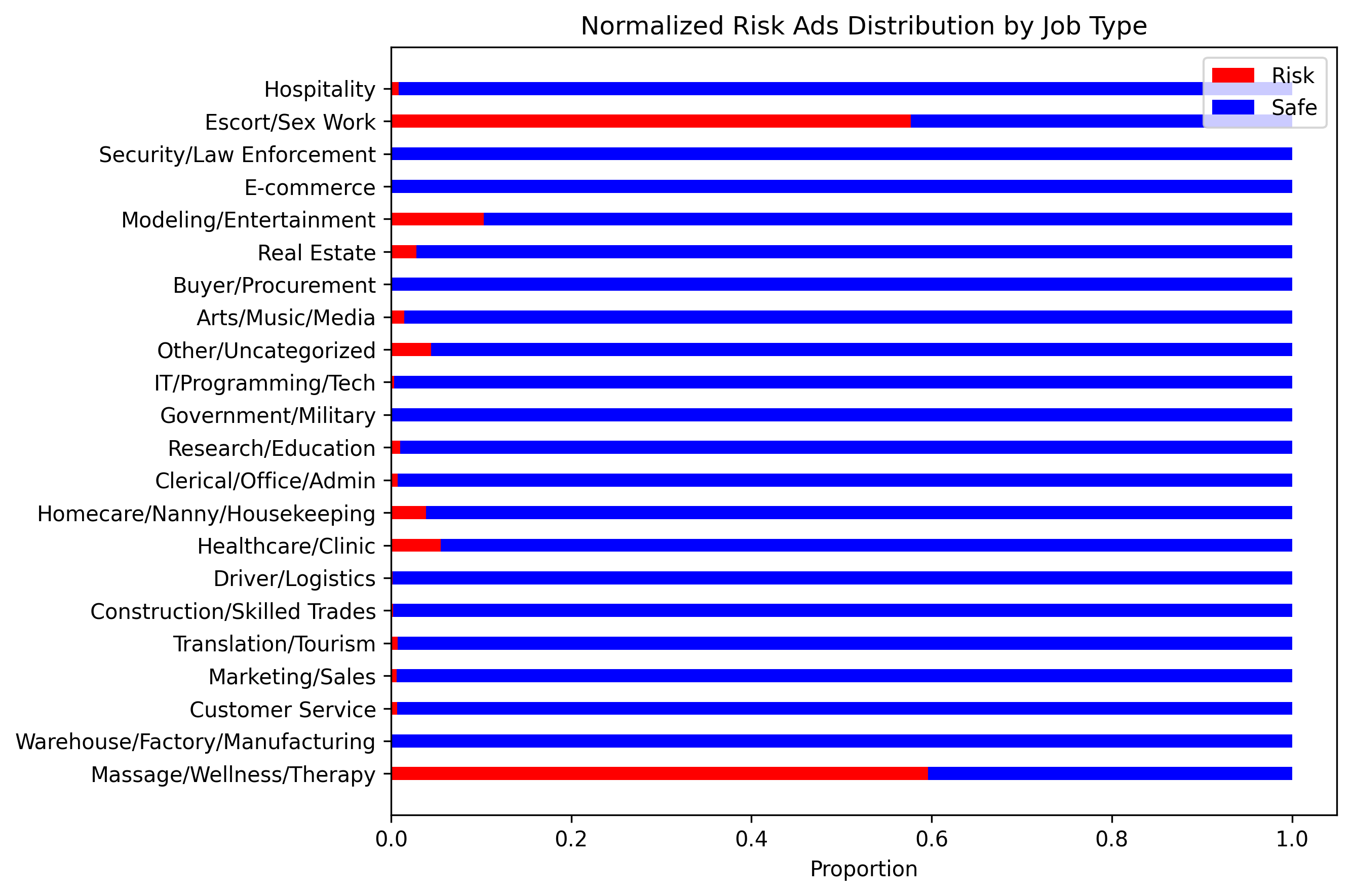}
  \caption{Distribution of industry in percentage}
  \label{fig:second}
\end{minipage}
\end{figure}

\subsection{Geographic, Industry, and Recruitment Characteristics of Risky Job Advertisements}

We first examine the geographic distribution of risky job advertisements based on the claimed job locations. In terms of absolute volume, New York and California account for the largest numbers of risky postings. New York contains 2,779 risky advertisements out of 19,424 total job postings, while California contains 2,065 risky advertisements out of 70,843 total postings.

However, states with the highest proportions of risky job advertisements exhibit a markedly different pattern. Pennsylvania shows the highest risk concentration, with 66.8\% of postings classified as risky (423 out of 633), followed closely by Ohio at 66.7\% (104 out of 156). Several other states also show high proportions of risky advertisements, including Iowa (59/91), Oklahoma (19/30), Tennessee (57/93), Massachusetts (235/425), South Carolina (203/369), and Maine (7/13), each with more than 50

The two states with the highest absolute volumes of risky advertisements—California and New York—are home to major international metropolitan areas that often serve as primary destinations for immigrants. These states are also major economic powerhouses in the United States, acting as central hubs for entertainment, finance, technology, and tourism, and attracting large immigrant populations \cite{bureau_of_economic_analysis_gdp_2022,lee_what_2024,shorris_new_2025}.The combination of high volume but relatively low proportion of risky ads in these states suggests that potential risks may be less visible when embedded within large, diverse, and heterogeneous labor markets.  Interestingly, most of the states with higher probability of risky ads are the states surrounding those super states. In these surrounding states, labor markets appear more concentrated with risky postings, potentially reflecting lower participation by legitimate recruiters on non-mainstream job boards. As a result, job advertisements that list these surrounding states as work locations may warrant heightened scrutiny, as recruitment risks may be more salient and less obscured by large volumes of legitimate postings.

We further examine whether job advertisements exhibit consistent location signals across three sources: the domain on which the ad is posted, the location the ad claims the job will be based in, and the geographic location associated with the provided phone number. Combining these three dimensions yields twelve possible location-matching scenarios, including cases in which all locations match or mismatch, cases in which one location mismatches while the other two align, and cases in which one location is unspecified or unknown while the remaining two either match or mismatch. Advertisements with two or more unspecified or unknown locations are classified as not comparable.

\begin{figure*}[t]
    \centering
    \includegraphics[width=\textwidth]{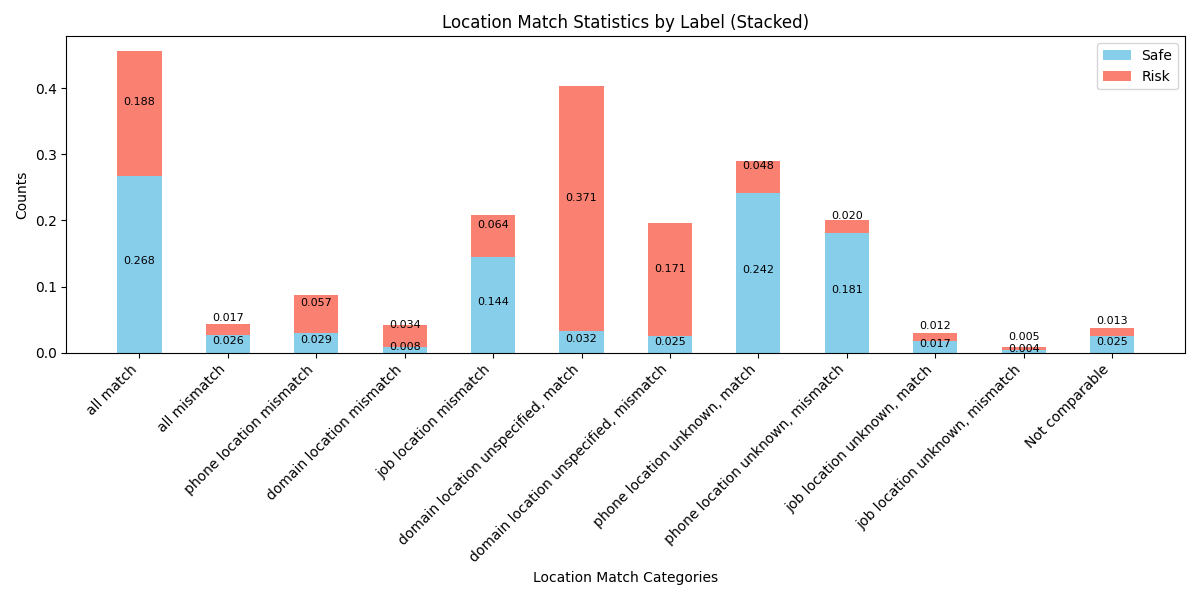}
    \caption{Percentage of post in different mismatch scenarios}
    \label{location_match}
\end{figure*}

Shown in Figure~\ref{location_match} and Table~\ref{tab:location_match_transposed_percent}, Risky ads are less likely than safe ads to exhibit fully consistent location information across all three sources (risk: 18.8\%; safe: 26.8\%). The largest divergence between safe and risky ads occurs in scenarios where the domain location and phone location are unspecified: risky ads show their highest proportions in categories where domain location is unspecified (37.1\% and 17.1\% for match and mismatch respectively), whereas safe ads are most concentrated in scenarios where the phone location is unknown or unspecified (24.2\% and 18.1\% for match and mismatch respectively). These patterns suggest that simple location consistency alone is not a strong indicator of risk. Instead, the platform on which an ad is posted and the type of contact information it provides appear to be more informative signals.

In particular, risky ads are more likely to appear on non-mainstream job boards that do not specify regional information, obscuring the ad’s geographic context. By contrast, manual inspection indicates that many safe ads lack phone-based location because they do not include phone numbers at all, relying instead on more formal contact channels. This distinction suggests that risky and safe ads differ less in whether locations match than in how transparently they disclose platform and contact information. 

We next analyze job industry categories. The massage business emerges as the most heavily trafficked industry in terms of both volume and proportion, with 9,558 risky advertisements out of 16,041 total postings. Uncategorized businesses (466 out of 10,532) and clerical/office positions (262 out of 37,246) also show notable volumes of risky advertisements. Closer inspection of the uncategorized category reveals that many of these advertisements recruit for highly specialized or ambiguous roles, such as personal accompaniment, which complicates standard occupational classification.

Importantly, while the uncategorized and clerical categories exhibit moderate volumes, their risk \emph{proportions} remain relatively low. In contrast, modeling-related advertisements show a higher concentration of risk, with 43 out of 417 postings (10.3\%) labeled as risky. This pattern suggests that many modeling calls may implicitly involve non-modeling labor functions, signaling elevated vulnerability despite lower total volume.

We exclude the sex work industry from further analysis because, although many postings in this category are labeled as risky, the job functions are typically explicit and consistent with the advertised labor. This makes deceptive recruitment and entrapment—our primary analytical focus—more difficult to identify using the current classification framework.

We further examine preferred gender requirements in risky job advertisements. In terms of raw volume, advertisements specifying a preference for female workers (5,654 out of 12,656) and those indicating no gender preference (4,695 out of 133,862) account for the largest number of risky postings. However, advertisements explicitly requesting female workers and couples exhibit substantially higher \emph{risk proportions}, with 44.6\% of female-only job ads and 30.7\% of couple-based job ads (260 out of 846) labeled as risky. This suggests that recruiters targeting specific relational or gendered labor configurations face elevated trafficking risk.

Finally, we examine preferred contact methods. Phone-based contact emerges as the most strongly associated with risky recruitment, with 10,132 risky advertisements out of 83,615 total postings using phone as the primary contact method (12.1\%). This pattern highlights direct voice or SMS communication as a dominant operational channel for trafficking-related recruitment.

\section{Discussion}

This study introduces a scalable, multilingual framework for detecting human trafficking–at-risk job advertisements using embedding models, supervised classifiers, and ensemble learning. Our results show that trafficking-related recruitment language exhibits consistent patterns across English, Chinese, and Russian contexts, enabling separation from many legitimate job advertisements in latent semantic space. PCA projections reveal relatively stable clustering between safe and risky posts, suggesting that trafficking-related content is linguistically distinguishable, though not perfectly separable .

Beyond classification performance, our analysis identifies systematic patterns in geographic, occupational, gender, and contact-method dimensions of risk. Risky recruitment activity is more prevalent in major economic hubs and surrounding regions, which may reflect the intersection of trafficking and smuggling dynamics, where migrants in precarious conditions become vulnerable to exploitation \cite{laczko_developing_2003,zhang_beyond_2012,united_nations_office_on_drugs_and_crime_unodc_global_2024}. At the same time, dense immigrant labor markets—characterized by high turnover, information asymmetries, and reliance on informal recruitment—may create opportunities for exploitative actors to operate alongside legitimate postings \cite{menjivar_legal_2011,abrego_sacrificing_2014}.

Our findings also suggest that trafficking risk cannot be inferred solely from geographic inconsistencies within job advertisements. Instead, domains lacking clear location information are more frequently associated with risky recruitment, consistent with evidence that exploitative actors may minimize traceable signals to reduce detectability \cite{latonero_rise_2012,united_nations_office_on_drugs_and_crime_unodc_global_2024}. However, alternative explanations, such as platform norms or informal posting practices, remain possible.

From a policy perspective, these results highlight the importance of monitoring non-mainstream job boards, where moderation and verification mechanisms are limited and illicit activity may be displaced from more regulated platforms \cite{farrell_policing_2020,latonero_rise_2012}. Industry-, gender-, and contact-level signals further help prioritize attention within these environments. The massage sector remains consistently associated with higher risk, while modeling-related advertisements show elevated risk despite lower volume \cite{zhang_beyond_2012}. Gender targeting patterns indicate higher risk in female-only and couple-based recruitment, and phone-based contact methods are strongly associated with risky postings, reinforcing the role of off-platform communication in trafficking activities \cite{latonero_rise_2012,farrell_policing_2020}.

Taken together, our findings demonstrate that trafficking recruitment exhibits consistent, learnable linguistic and structural patterns at scale, enabling the development of precise automated detection systems. The ensemble-based \textit{Trafficker Classifier} provides a foundation for content moderation, investigative triage, and prevention-oriented monitoring. More broadly, this work highlights the value of integrating multilingual NLP, network analysis, and platform governance to study human trafficking as a computationally observable and socially embedded phenomenon.
\subsection{Ethical Considerations and Limitations}

Despite these contributions, several limitations and ethical considerations warrant careful attention. First, our labeling strategy relies on a proxy measure of trafficking risk based on cross-platform phone number usage. Although UNODC reports this operational pattern is used by law enforcement on a routine basis \cite{united_nations_office_on_drugs_and_crime_unodc_global_2024}, it may introduce some false positives. For example, legitimate businesses or individuals may reuse contact information across multiple domains without engaging in exploitative practices. As a result, model predictions should be interpreted as indicators of potential risk rather than confirmed cases of trafficking.\\ \newline Second, false positives may disproportionately affect already vulnerable or stigmatized industries (e.g., massage, modeling), raising concerns about reputational harm, over-surveillance, or unintended enforcement consequences. The use of automated detection systems in sensitive contexts therefore requires caution, transparency, and human oversight. We emphasize that all model-identified at-risk advertisements should be subject to manual review and contextual investigation before any enforcement or intervention decisions are made. \\ \newline Finally, while our models demonstrate consistent performance across languages and domains, they do not achieve perfect accuracy, and the underlying data remain highly imbalanced. Future work should incorporate additional validation strategies, including human annotation and collaboration with domain experts, to further assess the reliability and fairness of the proposed framework.

\subsection{Limitations}

Several limitations should be acknowledged. First, although our dataset is multilingual, it is not globally representative of all trafficking recruitment ecosystems. Platform coverage remains uneven across regions and may systematically underrepresent encrypted, invitation-only, or dark web recruitment channels. Second, while our ensemble classifier achieves strong performance, all models depend on labeled training data, which inevitably reflects the observable trafficker behavior in the listed industrkes. Third, our analysis focuses primarily on textual content and structured metadata. We do not directly model visual, financial, or real-time interaction signals that may further strengthen trafficking detection. Finally, while our findings identify correlational patterns, they do not establish causal mechanisms of trafficking coordination, platform evasion, or recruiter network structure.\\
\textbf {Dataset and Code:} \newline
https://github.com/serenalyoko/humantraffcking

\paragraph{Acknowledgements.} The authors are grateful to the Many Hopes foundation, a 501(c)3 organization that supports research to combat human trafficking.

\bibliography{references}
\section{Appendix}

\subsection{Detailed Data summary}
\begin{table}[h]
\centering
\caption{Summary of Job Posts by Domain/platforms}
\label{tab:domains}
\small
\begin{tabular}{lrr}
\toprule
\textbf{Domain} & \textbf{Safe} & \textbf{Risk} \\
\midrule
rusrek & 4,423 & 180 \\
chineseinla & 45,054 & 885 \\
500work & 7,535 & 1,765 \\
dechinaren & 1,383 & 57 \\
nychinaren & 14,984 & 206 \\
chineseinsfbay & 14,216 & 232 \\
seattlechinaren & 1,872 & 63 \\
chineseinflorida & 1,546 & 137 \\
chineseinatlanta & 1,973 & 67 \\
vegaschinaren & 220 & 20 \\
meiguogongzuo & 0 & 451 \\
zgw580 & 0 & 93 \\
usahuarenjie & 0 & 404 \\
chicagochinaren & 0 & 1 \\
jinbay & 0 & 61 \\
chineseinhouston & 0 & 4 \\
huarenbang & 0 & 259 \\
chineseinpa & 0 & 14 \\
168worker & 0 & 1,574 \\
chineseinny & 0 & 11 \\ 
\midrule
\textbf{Total} & \textbf{93,206} & \textbf{6,484} \\
\textbf{Unique Total} & \textbf{83,025} & \textbf{2,816} \\
\bottomrule
\end{tabular}
\end{table}

\subsection{Examples of risk and safe job ads}

\begin{tcolorbox}[
  title= Example of risk job ad,
  box align=top,
  colback=gray!5,
  colframe=black,
  fonttitle=\bfseries,
  boxrule=0.5pt,
  left=6pt,
  right=6pt,
  top=6pt,
  bottom=6pt
]
Title: Atlanta recruits masseuses with incomes exceeding \$10,000

Job Description: Atlanta old shop recruits experienced female masseuses, age 43 and below, monthly income 12000+, single room, high speed internet, contact number 7702413449

\end{tcolorbox}

\begin{tcolorbox}[
  title= Example of safe job ad,
  box align=top,
  colback=gray!5,
  colframe=black,
  fonttitle=\bfseries,
  boxrule=0.5pt,
  left=6pt,
  right=6pt,
  top=6pt,
  bottom=6pt
]
Title: Teso Life Paid annual leave, medical insurance, 401K performance bonus Urgent recruitment Beauty shopping guide

Job Description: Beauty consultant/skin care guide Job requirements: Interested in cosmetics and skin care products, working experience in the same industry, full-time fairies are preferred. 1. Have a good image 2. Responsibility and ambition  3. Have a good team spirit 4. Legal work status 5. Have good communication skills Salary: Basic salary starting from 16 (salary increase in two weeks) Management through train in the fastest 3 months + bonus + equity incentive + professional management training course + employee discount Sam, Human Resources Department Contact number: 3474293421 Please be sure to indicate that you saw it on the New York Chinese Information Network when contacting, thank you',

\end{tcolorbox}

\subsection{Few-Shot Prompting}

We evaluate language model performance using both zero-shot and few-shot prompting strategies. In the zero-shot setting, the model receives no explicit guidance regarding which job advertisements should be considered risky. In contrast, the few-shot prompts incorporate dataset-specific instructions derived from UNODC guidelines and include one example each of a safe and a risky job advertisement.

Across our experiments, few-shot prompting consistently yields higher classification accuracy and more stable reasoning patterns than zero-shot prompting. Accordingly, we report results based on the few-shot configuration in the main analysis.

\begin{tcolorbox}[
  title=Prompt Used for Zero-shot Job Risk Classification,
  box align=top,
  colback=gray!5,
  colframe=black,
  fonttitle=\bfseries,
  boxrule=0.5pt,
  left=6pt,
  right=6pt,
  top=6pt,
  bottom=6pt
]
You are given a job posting. Your task is to analyze it for potential signs of human trafficking or forced labor.

\textbf{Respond strictly in JSON format:}
\begin{verbatim}
{
  "label": "SAFE" or "RISKY",
  "reason": "<short, clear reason>"
}
\end{verbatim}
\end{tcolorbox}

\begin{tcolorbox}[
  title=Prompt Used for Few-shot Job Risk Classification,
  box align=top,
  colback=gray!5,
  colframe=black,
  fonttitle=\bfseries,
  boxrule=0.5pt,
  left=6pt,
  right=6pt,
  top=6pt,
  bottom=6pt
]
You are a professional job-risk analysis expert.
Your task is to evaluate whether a job posting is SAFE or RISKY based on its content.

Decide using these dataset-specific rules:

\textbf{Mark RISKY when at least TWO of the following hold:}
\begin{enumerate}
  \item No employer identity or professional channel; only phone/SMS/IM contact.
  \item Compensation is unclear or absent (e.g., discuss in person / phone / high salary without numbers).
  \item Duties are vague or non-standard (e.g., collaborate / migration / easy VISA/certificate).
  \item Problematic screening or unrealistic claims.
\end{enumerate}

\textbf{Mark SAFE when the post includes normal job details such as:}
\begin{itemize}
  \item Clear pay or pay range, and taxable employment (Must be able to file tax, has SSN, etc),
  \item Specific duties, requirements, or schedules,
  \item Legitimate contact or employer identity (company name, HR email, fax, or website).
\end{itemize}

Here is a an example of label: \\
\begin{verbatim}
{
  "Post": example
  "label": "SAFE" or "RISKY",
  "reason": "<short, clear reason>"
}
\end{verbatim}
If uncertain, decide based on the clarity of job details; only mark as \textbf{RISKY} when strong warning signs appear.
\end{tcolorbox}

\begin{tcolorbox}[
  title=Selected response for safe post,
  box align=top,
  colback=gray!5,
  colframe=black,
  fonttitle=\bfseries,
  boxrule=0.7pt,
  left=7pt,
  right=7pt,
  top=7pt,
  bottom=7pt
]
\textbf{Input:} Teso Life Paid annual leave, medical insurance, 401K performance bonus Urgent recruitment Beauty shopping guide. Beauty consultant/skin care guide Job requirements: Interested in cosmetics and skin care products, working experience in the same industry, full-time fairies are preferred. 1. Have a good image 2. Responsibility and ambition  3. Have a good team spirit 4. Legal work status 5. Have good communication skills Salary: Basic salary starting from 16 (salary increase in two weeks) Management through train in the fastest 3 months + bonus + equity incentive + professional management training course + employee discount Sam, Human Resources Department Contact number: 3474293421 Please be sure to indicate that you saw it on the New York Chinese Information Network when contacting, thank you',
---------------------------------------------------------------------------

\textbf{Output:}
\begin{small}
    
\begin{verbatim}
{
  "label": "SAFE" ,
  "reason": "The job posting provides 
  a phone number for contact, which 
  is common in legitimate job postings, 
  but it is important to verify the 
  legitimacy of the contact 
  information.
  - The job description includes a 
  requirement for  (legal work status), 
  which suggests compliance with legal 
  employment standards.
  - The posting offers benefits such 
  as paid vacation, medical insurance, 
  401K, performance bonuses, and 
  professional management training, 
  which are typical of legitimate 
  employment opportunities.
  - The salary starts at a base rate 
  with the potential for a raise 
  within two weeks, and there is a 
  clear path to management positions, 
  which indicates transparency 
  in career progression.
  - There is no mention of fees for 
  employment, withholding of personal 
  documents, or any other indicators 
  commonly associated with 
  human trafficking or forced labor."
}
\end{verbatim}
\end{small}
\end{tcolorbox}

\begin{tcolorbox}[
  title=Selected response for risky post,
  box align=top,
  colback=gray!5,
  colframe=black,
  fonttitle=\bfseries,
  boxrule=0.7pt,
  left=7pt,
  right=7pt,
  top=7pt,
  bottom=7pt
]
\textbf{Input:} Online Business Marketing - Earn $500-$900 Daily MRR+PLR [RU]
DISCLAIMER: PLEASE CHECK YOUR SPAM EMAILS. I WILL RESPOND WITHIN 24 HOURS. ? Discover Financial Freedom With Our Exclusive Wealth Building Program? This Is A Guaranteed, Proven Process That Works! Hello, Aspiring Enthusiasts! Do you dream of a life free from the constraints of a 9-5 job? Or are you struggling to find a job? Do you envision a future where financial worries are a thing of the past? If you are ready to take control of your life and your finances, I have something special for you! ? Introducing: The Freedom Machines Program As a first-generation child of immigrants, I understand the hard work it takes to build something meaningful from the ground up. Today, I have the privilege of living the life I have always dreamed of – a life of financial abundance and freedom. Now it’s your turn, and I’m here to guide you! CLICK HERE NOW www.mindmoneyabundance.com/
---------------------------------------------------------------------------

\textbf{Output: }

\begin{small}
\begin{verbatim}
{
  "label": "RISKY" ,
  "reason": "
  - The job posting promises 
  high daily earnings of $500 to $900 
  with minimal time investment 
  (2 hours a day) and no technical 
  knowledge required. This could be 
  an indicator of a "too good to be 
  true" offer, which is often 
  associated with fraudulent 
  schemes or exploitative practices.
- The emphasis on financial freedom 
and the lack of requirement for a 
resume or work experience can be 
a tactic to attract vulnerable 
individuals who are desperate for 
employment, which is a common strategy 
in human trafficking and forced labor 
recruitment.
- The requirement for a "small 
investment in yourself" suggests 
that individuals may need to pay 
upfront fees, which is a red flag for
potential scams or exploitative 
practices.
- The job description uses persuasive 
language to create a sense of urgency 
and exclusivity, such as "only 
accepting 120 passionate people" and 
"time to act is now," which can pressure
individuals into making hasty decisions 
without fully understanding the terms.
- The lack of specific details about the
nature of the work, employment type, and 
location further adds to the suspicion, 
as legitimate job postings typically 
provide clear information about the job
role and requirements.
"
}
\end{verbatim}
\end{small}
\end{tcolorbox}

\clearpage

\end{document}